\documentclass[doublecol]{epl2}
\usepackage{amsmath}
\usepackage{amssymb}

\title{Repulsive Fermi gases in a two-dimensional lattice with non-Abelian gauge fields}

\author{Yi-Xiang Wang \inst{1,2} \and Fuxiang Li \inst{3} }

\institute{
  \inst{1} School of Science, Jiangnan University, Wuxi 214122, China.\\
  \inst{2} Department of Physics and Astronomy, University of Pittsburgh, Pittsburgh, Pennsylvania 15260, USA\\
  \inst{3} Center for Nonlinear Studies and Theoretical Division, Los Alamos National Laboratory, Los Alamos, NM 87545 USA.
}

\abstract{
Motivated by the recent experiment realizing bidirectional spin-orbit coupled (SOC) Bose-Einstein condensates (BEC), we theoretically explore the properties of repulsive fermions in the two-dimensional (2D) optical lattice with such non-Abelian gauge fields.  Within the mean-field level, we find a novel phase of topological antiferromagnetic (TAFM) order which incorporates both the non-trivial topology due to spin-flip hopping and spontaneous symmetry breaking (SSB) for the in-plane spin order.  We argue that the appearance of such a phase is generic for repulsive fermions in Chern-bands achieved through SOC.  Our work paves the way for further studies of fermionic generalization of 2D non-Abelian SOC quantum gases.}

\begin{document}

\maketitle

\section{Introduction}

In the past few years, fascinating progresses have been made in studying topological matters in cold atoms.  Many topological models and phenomena that are difficult to access in solid materials have been realized in optical lattices, such as the Harper-Hofstadter model \cite{M.Aidelsburger1,H.Miyake} and the Haldane model \cite{G.Jotzu}.  In particular, the spin-orbit coupling (SOC) that links a particle's motion to its spin \cite{X.J.Liu1,X.J.Liu2} have been successfully realized and manipulated in cold atoms.  Early theories and experiments on this regard chiefly focused on the SOC along one direction \cite{Y.J.Lin,P.Wang,L.W.Cheuk}, see i.e.  Refs.~\cite{H.Zhai,V.Galitski,N.Goldman} for reviews. In a recent experiment conducted by the Pan's group~\cite{Z.Wu}, SOC along two directions have been successfully achieved for Bose-Einstein condensates (BEC) in optical  lattices with Raman-assisted tunneling. Compared with unidirectional SOC, which is an Abelian gauge field, the bidirectional SOC corresponds to a genuine non-Abelian gauge potential and cannot be gauged away. Theoretically, the key properties of such a BEC have been analyzed, in which the impacts due to the coupling to higher bands are highlighted \cite{J.S.Pan}.  Another work studied the interplay of SOC and higher orbital bands for interacting bosons, and novel features are found in the Bogliubove excitation including Dirac and topological phonons~\cite{Y.Q.Wang}.  The dramatic difference of Bose gases in such non-Abelian gauge fields from those in Abelian ones needs more investigations, and the studies of their fermionic counterparts are still absent.

In this work, we theoretically discuss the properties of fermionic atoms loaded to the lowest orbital band of a 2D optical lattice with SOC similar to that in Ref.~\cite{X.J.Liu2,Z.Wu}. Here, the on-site repulsive Hubbard interaction is considered. This is partly motivated by the recent experimental advances of cooling the repulsive Fermi gases to temperatures near the N\'{e}el antiferromagnetic transition~\cite{afm1,afm2,afm3}, where long-range correlations across the whole lattice have been observed~\cite{afm3}. Theoretically, it has also been demonstrated that SOC can drastically change the many-body physics due to the modified single-particle band structures. For example, in a 1D Fermi gas with infinite repulsive interactions, even a tiny SOC can completely change the ground state spin texture~\cite{X.Cui}. Our work serves to illustrate the effects of repulsive interactions on Fermi gases subject to bidirectional SOC in the mean field level. The mean-field method applied in this work has been demonstrated to be qualitatively effective in dealing with the 2D \cite{V.S.Arun,A.M.Cook,T.I.Vanhala,J.He1,J.He2,W.Zheng} and 3D \cite{Y.X.Wang,J.Liu,B.Roy} correlated fermion systems in that it can capture the different correlations with the change of parameters in a many-body system.  Such a treatment also provides a starting point for applying more sophisticated analysis in the future, such as quantum Monte Carlo~\cite{qmc1} or density matrix renormalization group analysis~\cite{dmrg1,dmrg2}. 

An important observation here is the emergence of topological antiferromagnetic (TAFM) phase induced by the 2D SOC.  The coexistence of the topological bands and spontaneous symmetry breaking (SSB) has been of great interest. Previous work has suggested that the modified lattice symmetry due to AFM patterns could lead to new topological classifications~\cite{topoAFM}. For example, for attractive Fermi gases in a Chern band (i.e. the Haldane model), interaction could drive the Chern insulator into a topological superconducting phase before finally entering the trivial insulating phase~\cite{shizhong}. Here, with repulsive interactions, we find parallel phenomena that an in-plane AFM phase serves as the topological symmetry breaking phase during the interaction-driven topological phase transitions.

The basic picture of the TAFM phase can be understood as follows. The in-plane AFM orders together with the uniform tilting of spins along perpendicular direction induced by Zeeman fields give rise to both anti-parallel (${\cal M}_\perp$) and parallel (${\cal M}_z$) components for spin patterns in neighboring sites.  Since hoppings between the same/different spins involve trivial/non-trivial Peierls phases, ${\cal M}_z, {\cal M}_\perp$ together give rise to the non-trivial $\pm\frac{\pi}{2}$-flux when atoms are hopping around a plaquette, see Fig.~\ref{fig:hop} for illustrations.  When further increasing the interactions or the external Zeeman fields perpendicular to the plane, the in-plane AFM orders disappear and the spins are all polarized (${\cal M}_z\ne0,  {\cal M}_\perp=0$), giving rise to a topologically trivial phase where fermions gain a zero flux when hopping around a plaquette.

\section{Noninteracting Model}

We start with the ultracold fermions, such as $^{40}$K or $^6$Li atoms, trapped in a 2D square optical lattice \cite{Z.Wu} with lattice potential
$V_l({\boldsymbol r})=V_0[\text{cos}^2(k_0x)+\text{cos}^2(k_0y)]$,
and the commensurate Raman potential for SOC
$V_R({\boldsymbol r})=V_1[\text{cos}(k_0x)\text{sin}(k_0y)\sigma_x +\text{cos}(k_0y)\text{sin}(k_0x)\sigma_y]$.
Here $\boldsymbol{\sigma}$'s are Pauli matrices acting on the space of hyperfine atomic states; $k_0=\frac{2\pi}{\lambda_L}$ is the wave vector; $V_0$ and $V_1$ are the lattice depth and Raman coupling strength, respectively.   

\begin{figure}
\includegraphics[width=7.6cm]{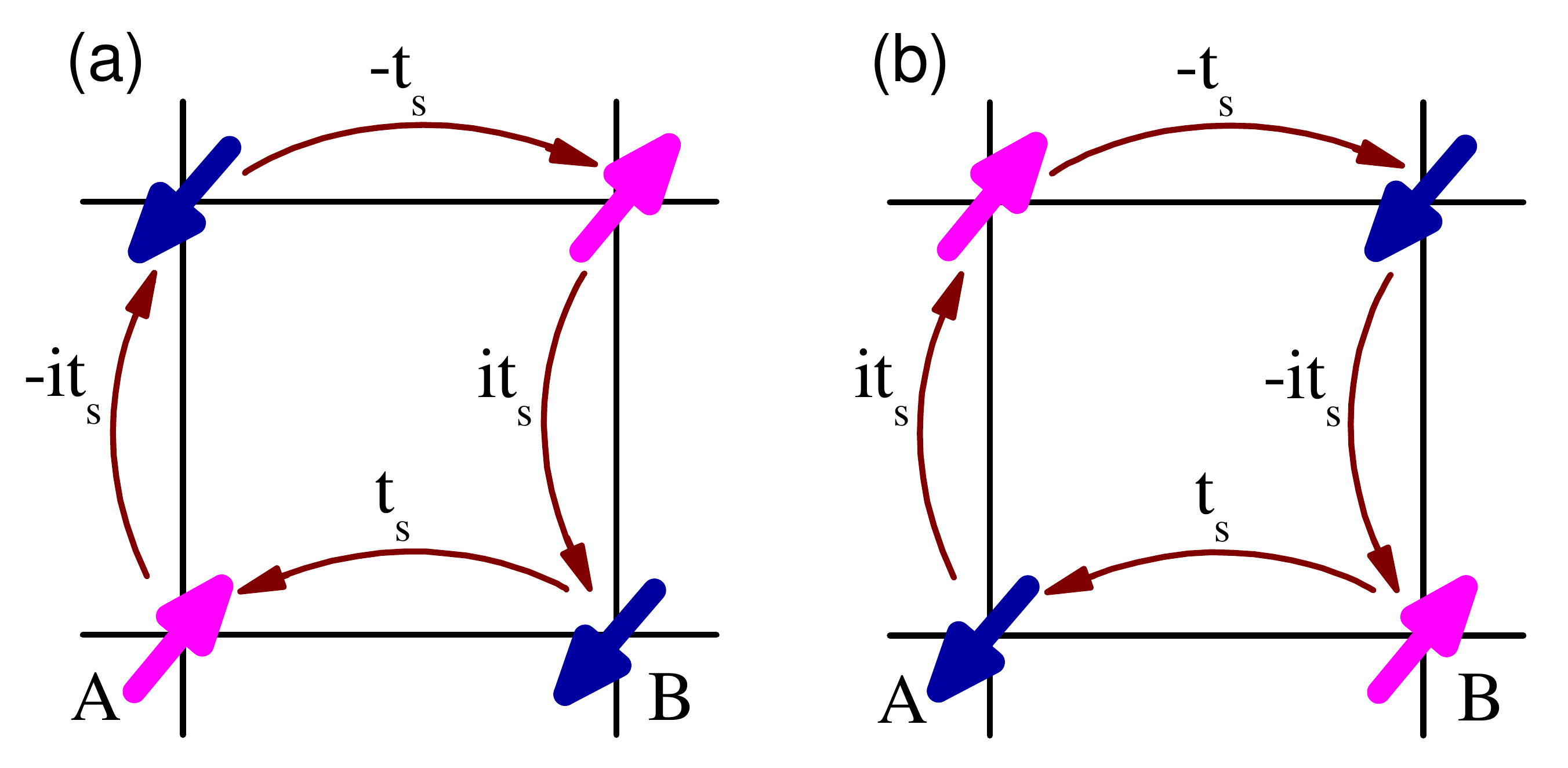}
\caption{(Color online) Schematic plot of Raman-assisted spin-flip hopping within a plaquette, where the amplitudes are set to be equal $t_{\text{SOx}}=t_{\text{SOy}}=t_s$. Note that spin-flip hoppings alone produces flux 0, though they involve non-trivial Peierls phases. Thus, it is necessary that spin-flip and spin-conserved hopping (without Peierls phases) {\em both} exist to produce the non-trivial $\pi/2$-flux. }
	\label{fig:hop}
\end{figure}

Here we consider only the lowest $s$-orbital band as the higher bands have energy separations from $ s $-band much greater than all energy scales we consider~\cite{pethicksmith}.  The dynamics of fermionic atoms are described by the following tight-binding Hamiltonian \cite{Y.Xu}:
\begin{align} \label{hamreal}
H_0=&-\sum_{{\boldsymbol j}s}(t_x\hat c_{{\boldsymbol j}s}^\dagger c_{{\boldsymbol j}+{\boldsymbol e}_x,s}
+t_y\hat c_{{\boldsymbol j}s}^\dagger c_{{\boldsymbol j}+{\boldsymbol e}_y,s}+\text{H.c.})
\nonumber\\
&+\sum_{\boldsymbol j}(-1)^{j_x+j_y}
\Big[t_{\text{SOx}} (\hat c^\dagger_{{\boldsymbol j}\uparrow}\hat c_{{\boldsymbol j}+{\boldsymbol e}_x,\downarrow}-c^\dagger_{{\boldsymbol j}\uparrow}\hat c_{{\boldsymbol j}-{\boldsymbol e}_x,\downarrow})
\nonumber\\
&+i t_{\text{SOy}} (\hat c^\dagger_{{\boldsymbol j}\uparrow}\hat c_{{\boldsymbol j}+{\boldsymbol e}_x,\downarrow}-\hat c^\dagger_{{\boldsymbol j}\uparrow}\hat c_{{\boldsymbol j}-{\boldsymbol e}_y,\downarrow})+\text{H.c.}\Big]
\nonumber\\
&+\sum_{\boldsymbol j}h_z(c_{{\boldsymbol j}\uparrow}^\dagger c_{{\boldsymbol j}\uparrow}
-c_{{\boldsymbol j}\downarrow}^\dagger c_{{\boldsymbol j}\downarrow}),
\end{align}
here $c_{{\boldsymbol j}s}$ ($c_{{\boldsymbol j}s}^\dagger$) is the annihilation (creation)
operator for a fermion at lattice site $\boldsymbol j$, with spin $s=\uparrow,\downarrow$; $t_{x(y)}$ denote the spin-conserved nearest-neighbor (NN) hoppings and $t_{\text{SOx}(\text{SOy})}$ are the spin-flipping NN hoppings due to SOC; $h_z$ is the  linear Zeeman term. See Fig.~\ref{fig:hop} for illustrations of the spin-flip hoppings. Due to the SOC, the unit cells are doubled and we have a checkerboard pattern of alternating A-B sublattices along both $x$ and $y$ directions.

Transforming to the momentum space and choosing the basis $\Psi({\boldsymbol k})=(c_{A{\boldsymbol k}\uparrow},c_{A{\boldsymbol k}\downarrow},e^{ik_xa}c_{B{\boldsymbol k}\uparrow},e^{ik_xa}c_{B{\boldsymbol k}\downarrow})^T$, one can write the Hamiltonian as $H_0=\sum_{{\boldsymbol k}}\Psi({\boldsymbol k})^\dagger h_0({\boldsymbol k})\Psi({\boldsymbol k}) $, with
\begin{align}\label{h0k}
h_0({\boldsymbol k})=-h_t\tau_x+(d_x\sigma_x-d_y\sigma_y)\tau_y +h_z\sigma_z,
\end{align}
here $\boldsymbol{\sigma}$ and $\boldsymbol{\tau}$ are Pauli matrices acting on the spin and sublattice spaces respectively, and
$h_t=2[t_x\text{cos}(k_x a)+t_y\text{cos}(k_y a)]$,
$d_x=2t_{\text{SOx}}\text{sin}(k_xa)$,
$d_y=2t_{\text{SOy}}\text{sin}(k_ya)$.
For simplicity, the parameters are chosen to be uniform $t_x=t_y=t$, $t_{\text{SOx}}=t_{\text{SOy}}=t_s$, and $t=1$ is taken as the unit of energy in the following.    The Hamiltonian $h_0({\boldsymbol k})$ possesses the inversion symmetry as ${\cal I} h_0({\boldsymbol k}){\cal I}=h_0(-{\boldsymbol k})$, with ${\cal I}=\tau_x$.  When the Zeeman field is absent $h_z=0$, one can further define the time-reversal symmetry (TRS) for the spin-orbit coupled system,  ${\cal T}h_0({\boldsymbol k}){\cal T}^{-1}=h_0(-{\boldsymbol k})$, and the particle-hole symmetry ${\cal C}h_0({\boldsymbol k}) {\cal C}^{-1}=-h_0(-{\boldsymbol k})$, with ${\cal T}=i\sigma_y\tau_x{\cal K}$ and ${\cal C}=\tau_y$. Here $\cal K$ is the complex conjugation. In the presence of $h_z$, the symmetries ${\cal T}, {\cal C}$ and their combination ${\cal TC}$ are all broken. Thus, generally, the system belongs to class A in the Altland and Zirnbauer notations and is characterized by the Chern number in 2D \cite{kanemele,shinsei}.

\begin{figure}
\includegraphics[width=8cm]{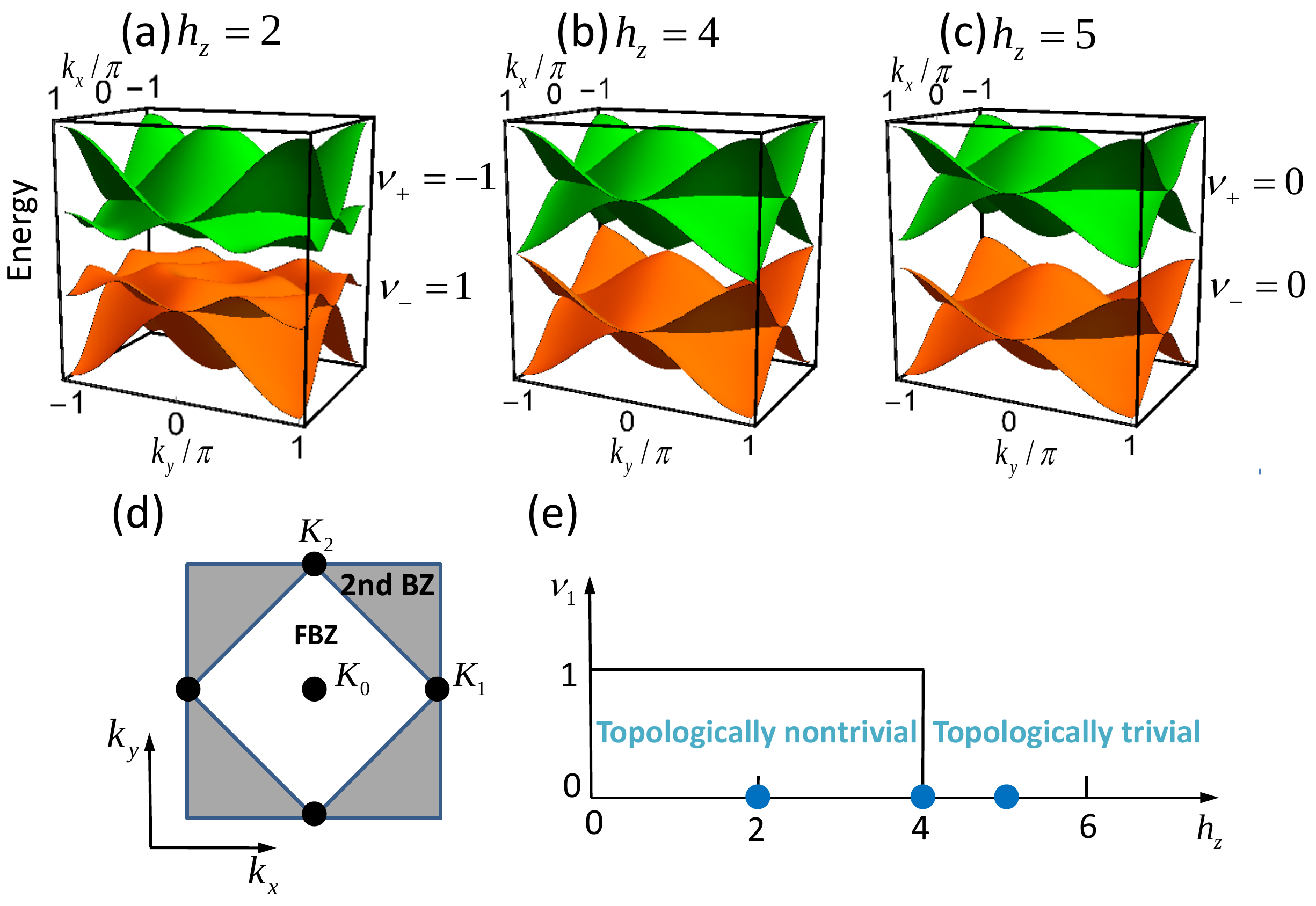}
\caption{(Color online) Energy spectrum and the topology of the Bloch bands of the noninteracting model in Eq.~(\ref{hamreal}). Here $t_x=t_y=1$ and $t_{\text{SOx}}=t_{\text{SOy}}=0.48$.  (a)-(c) The band dispersions with different Zeeman field $h_z$. The topological numbers $\nu_{+/-}$ are computed for the total upper/lower superbands.  (d) The resulting BZ and the positions of the Dirac points for the higher two superbands (green) to touch the lower two (orange). When $h_z=\pm4$, a Dirac corn forms at ${\boldsymbol K}_0=(0,0)$, as in (b); when $ h_z=0 $, Dirac corns form at ${\boldsymbol K}_{1,2}$.  (e) The Chern number $\nu_1$ for lower bands during the change of $h_z$.}
\label{fig:bands}
\end{figure}

The Hamiltonian can be brought into a block-diagonal form which greatly facilitates the following analysis. To do this, we first rearrange the basis $\Psi({\boldsymbol k})$ as $\Psi'({\boldsymbol k})=(c_{A{\boldsymbol k}\uparrow},e^{ik_xa}c_{B{\boldsymbol k}\uparrow},c_{A{\boldsymbol k}\downarrow},e^{ik_xa}c_{B{\boldsymbol k}\downarrow})^T$, and then make unitary transformation with the matrix
\begin{align}\label{unitaryU}
U=\frac{1}{\sqrt2}\begin{pmatrix}
1& 0& -1& 0
\\
1& 0& 1& 0
\\
0& 1& 0& 1
\\
0& -1& 0& 1
\end{pmatrix}.
\end{align}
The resulting Hamiltonian $h_0'({\boldsymbol k})=U^\dagger h_0({\boldsymbol k})U$ then reads
\begin{align} \label{blockh0k}
h_0'({\boldsymbol k})= 
\begin{pmatrix}
g_{1{\boldsymbol k}} & 0 \\ 0 & g_{2{\boldsymbol k}}
\end{pmatrix}, 
\end{align} 
with $g_{\gamma{\boldsymbol k}}=[h_z-(-1)^\gamma h_t]\tau_z-d_y\tau_x-d_x\tau_y$, where $\gamma=1,2$ denotes the two blocks $g_{1,2{\boldsymbol k}}$.  From this form, one can directly obtain the spectrum as $\varepsilon_{\gamma \pm}({\boldsymbol k})=\pm\sqrt{d_x^2+d_y^2+[h_t-(-1)^\gamma h_z]^2}$. Further, in the boundary of the first Brillouin zone (FBZ), $k_x=\pm k_y\pm\pi $, $g_{1\boldsymbol k}=g_{2,\boldsymbol k}=h_z\tau_z-d_y\tau_x-d_x\tau_y$, and therefore the two ``upper bands" $\varepsilon_{1,2 +}$ will be degenerate, as well as for the two ``lower bands" $\varepsilon_{1,2 -}$.  (See Fig. \ref{fig:bands} (a-c) where green bands are $\varepsilon_{1,2+} $ and orange bands are $\varepsilon_{1,2-}$).  Such a degeneracy, however, is a consequence of the fine-tuned parameters $t_x=t_y=1$.  Upon increasing anisotropies of $t_x\ne t_y$, such a degeneracy will be lifted and the two bands of the same color in Fig.~\ref{fig:bands} only touch at discrete points given by $h_t({\boldsymbol k})=0$. In the following, we stick to the $t_x=t_y=1$ limit as is typically the case in experiments. We have checked that the following results are unaffected by such an accidental degeneracy.

Now we consider the topological properties of the model. Since in the isotropic limit the upper/lower two bands are pairwise-connected respectively, we refer them as two ``superbands" (represented by two colors in Fig.~\ref{fig:bands}(a-c)). The two superbands touch each other when $h_z=0, \pm4$ where Dirac cones form at the FBZ center ${\boldsymbol K}_0=(0,0)$ or corners ${\boldsymbol K}_1=(0,\pi), {\boldsymbol K}_2=(0,\pi)$ (see Fig. \ref{fig:bands} (d)).  The total Chern numbers for the two superbands are
\begin{align} \label{chernformula}
\nu_\pm=\nu_{1\pm}+\nu_{2\pm}, \quad 
\nu_{\gamma\pm}=\pm\int\frac{d^2\boldsymbol k}{4\pi} \hat{g}_{\gamma\boldsymbol k} \cdot (\hat{g}_{\gamma\boldsymbol k}\times \hat{g}_{\gamma\boldsymbol k}),
\end{align}
where $\hat{g}_{\gamma \boldsymbol k} = g_{\gamma \boldsymbol k}/|g_{\gamma \boldsymbol k}|$. In fact, the topological phases can be more easily identified by checking the change of sign of mass term when expanding $g_{1,2\boldsymbol k}$ around the Dirac points near phase transitions. When $|h_z|\rightarrow\infty$, $\hat{g}_{1,2\boldsymbol k}$ tends to point towards north/south poles and the Chern number is trivial $\nu_\pm=\nu_{1\pm}=\nu_{2\pm}=0$. The mass term for $g_{1\boldsymbol k}$ ($g_{2,\boldsymbol k}$) changes sign at $h_z=0,-4$ ($h_z=0,+4$); thus, the topologically non-trivial phase occurs at $h\in(-4,4)$. Explicit calculations show that in $h_z\in (0,4]$, $\nu_{1-}=0, \nu_{2-}=+1$, while for $h_z\in[-4,0)$, $\nu_{1-}=-1, \nu_{2-}=0$. In summary, we have the total Chern number for superbands
\begin{align}
\nu_-=-\nu_+=\text{sgn}(h_z)-\frac{1}{2}[\text{sgn}(h_z-4)+\text{sgn}(h_z+4)].
\end{align}
The phase diagram for the half-filled situation is shown in Fig.~\ref{fig:bands}(e) and the topological phase transition is clearly controlled by the Zeeman field.

\section{Hubbard interaction}

For fermions in optical lattices subject to short-range interactions, the system can be described by the Fermi-Hubbard model \cite{pethicksmith,Esslinger2010}
$H=H_0+H_I$, where $H_0$ is given previously in Eq.~(\ref{hamreal}), and the onsite interaction is
\begin{eqnarray}
H_I=U\sum_{{\boldsymbol j}\alpha}\hat n_{{\boldsymbol j}\alpha\uparrow}\hat n_{{\boldsymbol j}\alpha\downarrow}.
\end{eqnarray}
Here ${\boldsymbol j}$ denotes the unit cell, $\alpha=A,B$ the sublattices, $U>0$ the repulsive interaction strength and $\hat n_{{\boldsymbol j}\alpha s}=\hat c_{{\boldsymbol j}\alpha s}^\dagger c_{{\boldsymbol j}\alpha s}$ the fermionic number operator.  We consider the half-filling case, i.e., the number of fermionic atoms being equal to the number of lattice sites.  For repulsive interactions, the relevant channels are charge/spin density waves, which are captured by the order parameters~\cite{V.S.Arun,Y.X.Wang}
\begin{align} \label{rhomf}
\rho_{{\boldsymbol j}\alpha} &= \sum_s\langle n_{{\boldsymbol j}\alpha s}\rangle,
\\\label{mmf}
{\bm{\mathcal M}}_{{\boldsymbol j}\alpha} &= \sum_{s,s'}\langle c^\dagger_{{\boldsymbol j}\alpha s}{\boldsymbol\sigma}_{ss'}c_{{\boldsymbol j}\alpha s'}\rangle,
\end{align}
where $\langle\cdots\rangle$ denotes average taken at the ground state or a thermal ensemble. We look for the solutions invariant under unit-cell translations, i.e., $\rho_{{\boldsymbol j}\alpha}=\rho_\alpha$ and ${\boldsymbol{\mathcal M}}_{{\boldsymbol j}\alpha} = \bm{\mathcal{M}}_\alpha$ are independent of $\boldsymbol j$.  Then the Hubbard interaction can be decoupled as \cite{A.M.Cook,Y.X.Wang}
\begin{align}
&&H_I^d=\frac{U}{2}\sum_{{\boldsymbol j}\alpha s}\rho_{\alpha}\hat c_{{\boldsymbol j}\alpha s}^\dagger \hat c_{{\boldsymbol j}\alpha s}-\frac{U}{2}\sum_{{\boldsymbol j}\alpha ss'} \hat c_{{\boldsymbol j}\alpha s}^\dagger {\bm{\mathcal M}}_{\alpha}\cdot {\boldsymbol \sigma}_{ss'} \hat c_{{\boldsymbol j}\alpha s'}
\nonumber\\ \label{hammf}
&&-UN_c\sum_\alpha\langle n_{\alpha\uparrow}\rangle \langle n_{\alpha\downarrow}\rangle
+UN_c\sum_\alpha\langle{\cal M}_{\alpha+}\rangle
\langle{\cal M}_{\alpha-}\rangle.
\end{align}
where ${\cal M}_{\alpha\pm}={\cal M}_{\alpha x}\pm i{\cal M}_{\alpha y}$.
Here the first line represents the decoupled single-particle terms while the second line denotes the constant terms for a given configuration of cells, with $N_c$ the cell number.  Although the constant terms do not affect the topological property of the system, they must be included in calculating the total energy.  The order parameters $ \rho_\alpha, {\bm{\mathcal M}}_\alpha $ are obtained by iteratively solving for the self-consistent conditions in Eqs.~(\ref{rhomf})--(\ref{hammf}).  We choose different initial states to ensure that the converged ground state owns the lowest energy.  In our calculation, the size for the system is taken as $L_x=L_y=40$ and the periodic boundary condition is applied.

Explicit calculations show that $\rho_{A}=\rho_{B}=1$, as is expected.  Because at half filling, the system is a band insulator and already possesses a charge-gap (except at $h_z=0,\pm4$) and the charge density waves cannot lower the ground state energy through gap-openning. However, the spins are still free to rearrange themselves because the lower superbands involves a mixture of both spins.  

In the mean field level, a uniform ${\cal M}_{\alpha z}$ corresponds to a renormalization of Zeeman fields $h_z$, leading to the shift of the quantum critical points for topological transitions. Meanwhile, ${\cal M}_z$ can enhance real hopping between the same spins, while the in-plane ordering ${\cal M}_{\alpha x,y}$, if in an alternating AFM pattern, enhances the SOC-type of hopping in Fig.~\ref{fig:hop}, which carries Peierls phases.

\section{Spontaneous symmetry breaking}

\begin{figure}[h]
\includegraphics[width=8.8cm]{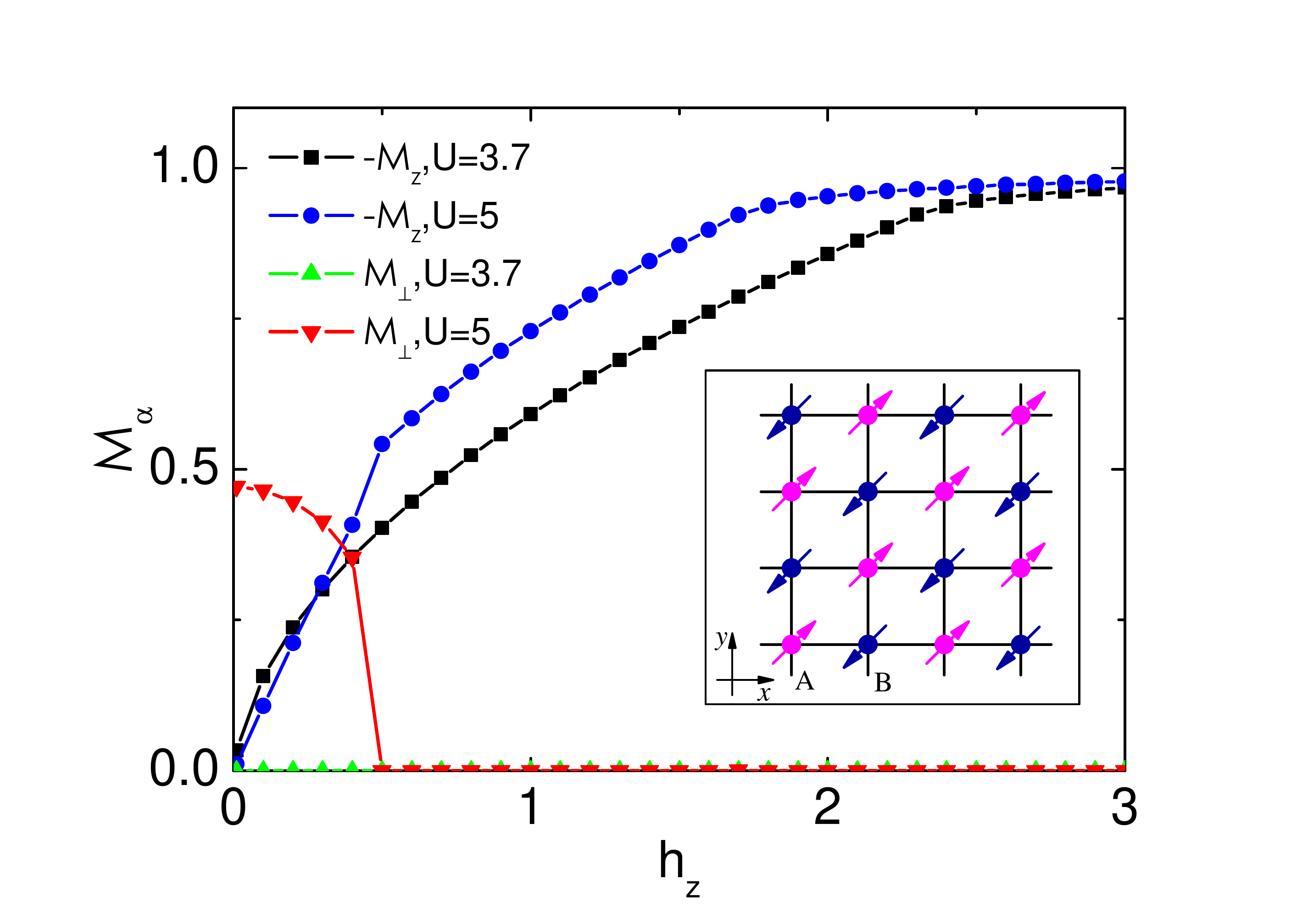}
\caption{(Color online) Magnetization ${\cal M}_\alpha$ versus the Zeeman energy $h_z$ at $ T=0 $ for different Hubbard $U$. Here $t_s=0.48$. The AFM order${\cal M}_\perp\neq0$ occurs for $|h_z|<0.5$.  Insets: In-plane spin patterns of AFM order.} \label{fig:zeroT}
\end{figure}

We first discuss the magnetization ${\cal \boldsymbol M}_\alpha$ obtained in the self-consistent calculations at zero temperature $ T=0 $. The perpendicular magnetization ${\cal M}_z$ turns out to have uniform magnitudes and directions in both sublattices
\begin{align}
{\cal M}_{Az}={\cal M}_{Bz}={\cal M}_z,
\end{align}
pointing opposite to the direction of external field $h_z$ (due to the sign choice in front of $ h_z $ in Eq.~(\ref{hamreal})). This should be viewed rather as the usual Zeeman effect in a paramagnet because the Ising symmetry for up and down spins is already broken by $h_z$. Also, the strength of ${\cal M}_z$ is proportional to that of the field $h_z$, as can be seen in Fig.~\ref{fig:zeroT}. The different interaction strength gives different spin susceptibility and therefore different response of ${\cal M}_z$ to $h_z$. On the other hand, the O(2) spin rotation symmetry within the lattice plane is preserved without interaction. We find that with interaction, if there is a SSB, it occurs in the pattern
\begin{align}
{\cal M}_{Ax}={\cal M}_{Ay}=-{\cal M}_{Bx}=-{\cal M}_{By}={\cal M}_\perp.
\end{align}
That is, the AFM order forms a stripe pattern parallel to the diagonal direction of the square lattice, with alternating signs in A, B sublattices. (See the inset of Fig.~\ref{fig:zeroT}).  Such an AFM order can also be considered as the spin density waves with the specific wave vector of ${\boldsymbol{Q}=(\pi,\pi)}$.  Compared with Fig.~\ref{fig:hop}, we see that such an AFM pattern amounts to choosing a given direction for the spin-flip hopping due to the spontaneous breaking of O(2) rotation symmetry.  The important thing is that such a pattern still allows for the flux-carrying hopping as shown in Fig.~\ref{fig:hop}, which is the basis for the TAFM phase.

The magnetizations ${\cal M}_{z}, {\cal M}_\perp$ at $T=0$ are shown in Fig.~ \ref{fig:zeroT} with the change of Zeeman fields $h_z$ and for two interaction strengths $U$.  As explained before, the spin susceptibility along $z$-direction increases with $U$, and therefore ${\cal M}_z$ saturates to the maximum value $-1$ faster for $U=5$ than for $U=3.7$.  On the other hand, the in-plane magnetization ${\cal M}_{\perp}$ does show the signature of SSB: for small interaction $U=3.7$, ${\cal M}_\perp=0$. This is because of the vanishing density of states, the Dirac points are stable against weak interactions. While for larger $U=5$, ${\cal M}_\perp\ne0$ provided the Zeeman field is not too strong.  The sign of ${\cal M}_\perp$ is fixed by the direction of $h_z$.  Further, the total magnetization satisfies $\bm{\mathcal M}^2={\cal M}_z^2+2{\cal M}_\perp^2 \le 1 $ due to the one-particle-per-site constraint, implying that with the increase of spin polarization along $z$ by $h_z$, i.e. ${\cal M}_z|_{h_z\rightarrow\infty}\rightarrow \pm1$, the in-plane AFM will be destroyed.  Such a physical picture is fully confirmed by the numerical results shown in Fig.~\ref{fig:zeroT}, where the in-plane AFM order sets in when $h_z<0.5$.  From the inset of Fig.~\ref{fig:zeroT}, we see that the lattice $C_4$ rotation symmetry is broken by the AFM phases. For this reason, some literatures also call such an ordering a "nematic phase" \cite{A.M.Cook}.  Here we choose to use the terminology AFM because the order parameter Eq.~(\ref{mmf}) is of the second order.

\section{Interacting phase diagram at $T=0$}

Now we turn our attention to the modified band structure and its topological properties due to $ {\cal M}_z, {\cal M}_\perp $.  Note that Eq.~(\ref{hammf}) only involves onsite terms, and therefore its effect is to renormalize the parameters of the non-interacting Hamiltonian Eq.~(\ref{hamreal}).  To see this, we transform Eq.~(\ref{hammf}) to momentum space using the basis in Eq.~(\ref{h0k}), which reads
\begin{align}
h_I({\boldsymbol k})=-\frac{U}{2}\left[ {\cal M}_z \sigma_z + {\cal M}_\perp (\sigma_x+\sigma_y)\tau_z \right].
\end{align}
One can similarly rearrange the basis and then perform a unitary transform to obtain the interaction-dressed Hamiltonian $h'({\boldsymbol k})=U^\dagger [h_0({\boldsymbol k}) + h_I({\boldsymbol k})] U$, with
\begin{align}
h'({\boldsymbol k}) &= \begin{pmatrix}
g'_{1\boldsymbol k} & 0\\ 0 & g'_{2\boldsymbol k}
\end{pmatrix}, 
\end{align}
with $g'_{\gamma\boldsymbol k}=[h_z'-(-1)^\gamma h_t]\tau_z-d'_{\gamma y}\tau_x-d'_{\gamma x}\tau_y$, $\gamma=1,2$, which is in completely the same form as the non-interacting one. The modified parameters are
\begin{align}
h'_z = h_z - \frac{U{\cal M}_z}{2}, \quad
d'_{\gamma x,y} = d_{x,y} - (-1)^\gamma \frac{U{\cal M}_\perp}{2}.
\end{align}
Note that ${\cal M}_z<0$ from Fig.~\ref{fig:zeroT}, the perpendicular magnetization serves to enhance the Zeeman field as expected.  Compared with the non-interacting vectors $d_x, d_y$ in Eq.~(\ref{h0k}), we see that the in-plane ones ${\cal M}_\perp$ break the inversion symmetry.  Then, we can again easily read out the interaction-dressed spectrum (up to some constants)
\begin{align}
\varepsilon_{\gamma\pm}({\boldsymbol k})=\pm\sqrt{d_{\gamma x}^{'2}+d_{\gamma y}^{'2}+[h_z'- (-1)^\gamma h_t]^2}. 
\end{align}
We see that the degeneracy within each superband is lifted from the whole BZ boundary to discrete points given by
\begin{align}
t_s{\cal M}_\perp(\text{sin}k_x+\text{sin}k_y)+(\frac{2h_z}{U}-{\cal M}_z)(\text{cos}k_x+\text{cos}k_y)=0.
\end{align}
Note that the above equation is always satisfied at the BZ corners ${\boldsymbol K}_1$ and ${\boldsymbol K}_2$, so the superband structure persists.

The in-plane AFM order competes with the SOC, as seen in $d'_{\gamma x,y}$, leading to the change of Dirac points.  If the condition $\frac{U{\cal M}_\perp}{4t_s}<1$ is satisfied, the original Dirac points will be moved or splitted, generating four new ones of $(X,X)$, $(-X,-X)$,  $(X,\pi-X)$ and $(\pi-X,X)$ in the BZ, with {\color{red} $X$=arcsin$\frac{U{\cal M}_\perp}{4t_s}$}.  The corresponding Chern number for the superbands can be calculated with the formula in Eq. (\ref{chernformula}) and is given as
\begin{align}
\nu_-=\text{sgn}(h_z')-\frac{1}{2}[\text{sgn}(h_z'+4\text{cos}X)
+\text{sgn}(h_z'-\text{cos}X)].
\end{align}
If $\frac{U{\cal M}_\perp}{4t_s}>1$, the Dirac points will be broken as $d_{\gamma x(y)}'$ cannot be vanishing in the whole BZ.  The phase diagram of interacting fermions when $t_s=0.48$ at zero temperature is given in Fig.~\ref{fig:topo}, in which four distinct phases appear, i.e., A: Chern insulator $(\nu_-=1,{\cal M}_\perp=0)$,
B: TAFM $(\nu_-=1,{\cal M}_\perp\neq0)$,
C: normal AFM $(\nu_-=0,{\cal M}_\perp\neq0)$,
and D: trivial insulator $(\nu_-=0,{\cal M}_\perp=0)$. We summarize the features of the phase diagram below:

\begin{figure}[h]
\includegraphics[width=8.8cm]{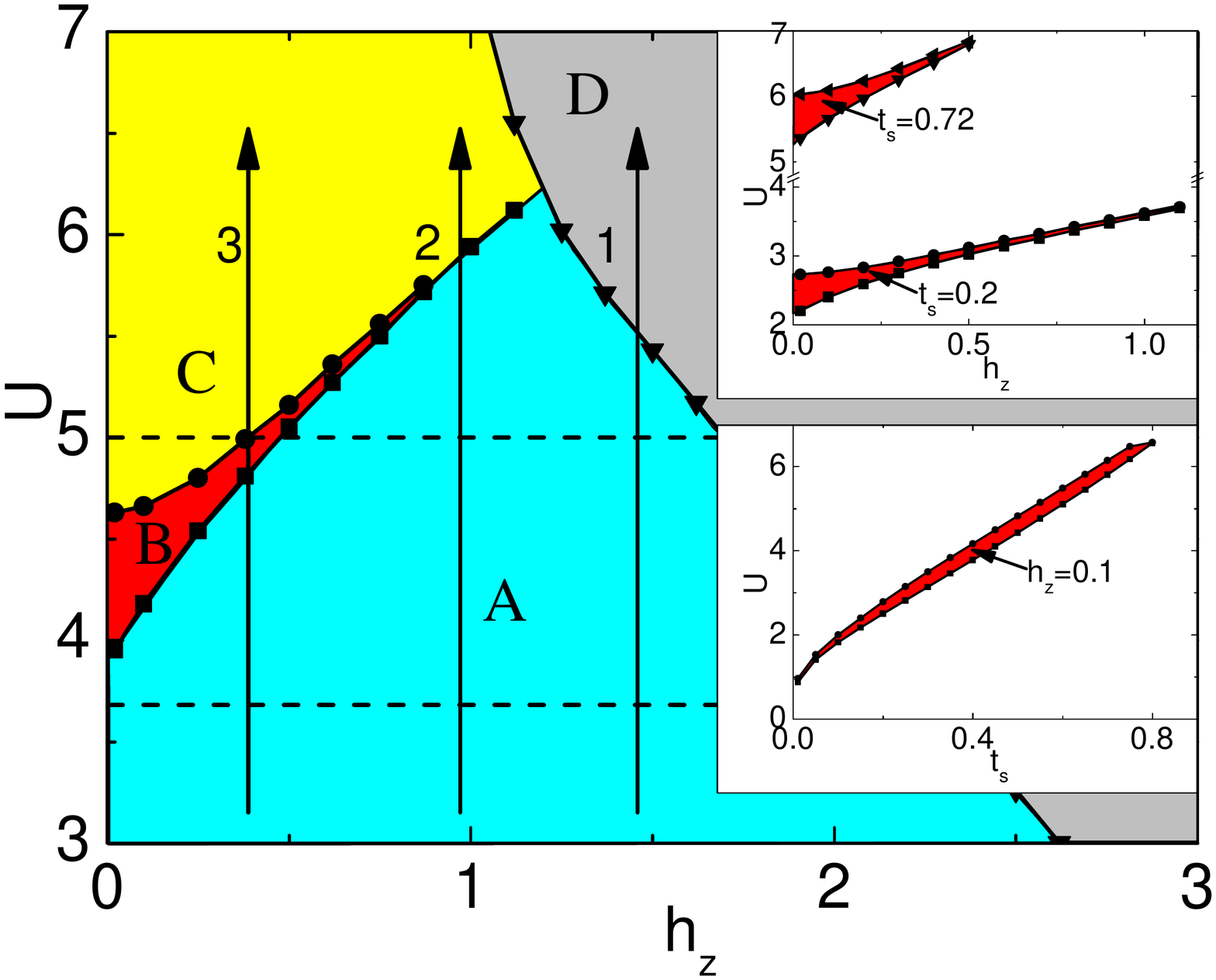}
\caption{(Color online) The zero-temperature phase diagram with parameters $(h_z,U)$.  There appear four phases of A: Chern insulator, B: TAFM, C: normal AMF and D: trivial insulator.  The TAFM phase carries Chern number $\nu_-=1$, representing a topological phase with SSB. The horizontal dashed lines show the $U$'s in Fig.~\ref{fig:zeroT}. $t_s=0.48$ is the same as in Fig.~\ref{fig:zeroT}.  The upper and lower insets show the spanned parametric regions of TAFM phase for certain $t_s$ and $h_z$, respectively.  \label{fig:topo}}
\end{figure}

i) Starting from the noninteracting case, when the Hubbard interaction $U$ increases, spin susceptibility along $z$-direction ramps up, leading to a larger effective Zeeman field which tends to shrink the topological non-trivial regions. For large external Zeeman field $h_z\in(1.07,4)$, as along the arrow 1, the symmetric Chern insulator phase is directly connected to the topological trivial phase via a gap closure.

ii) For smaller external fields $h_z$, however, the interaction leads to the second effect of inducing the SSB, i.e., in-plane AFM ordering ${\cal M}_\perp$.  When $h_z\in(0.87,1.07)$, as along arrow 2, if $U$ crosses the critical line, the SSB appears and the system enters a magnetic ordered phase. Our calculation shows that this phase has the Chern number of $\nu_-=0$, so it is the normal AFM phase, which interpolates between the symmetric Chern insulator and topological trivial phase.  Further increasing the interaction, the spin polarization along $ z $-direction finally prevails and diminishes the in-plane AFM order.  In this process, the bulk gap also closes and the system enters the symmetric topologically trivial phase.

iii) When the Zeeman field further decreases to $h_z<0.87$, as along arrow 3, the TAFM phase appears before entering the normal AFM phase. In the boundary between the symmetric Chern insulator and the TAFM phases, although the positions of Dirac points will change along with the SSB, the bulk gap does not close and thus the two phases are adiabatically connected.  In this sense, the phase is dubbed as TAFM.  As commented before, the TAFM spin pattern allows for spin-flip hoppings, which can carry a non-trivial Peierls phase.  The TAFM phase can accommodate the topological non-trivial phase as well as the SSB, as we verified.  Futher increasing the interaction, the bulk gap closes and correspondingly the Chern number changes from $\nu_-=1$ to $\nu_-=0$ and the system enters the normal AFM phase.  Finally,  the symmetric topologically trivial phase will dominate the system, just like the above case.

It is worthy pointing out as the finite Zeeman field $h_z>0$ breaks the TRS, both the symmetric and symmetric broken phases belong to class A in the topological classification \cite{kanemele,shinsei} and are characterized by the Chern number.  As a result, it leads to the observation of C=1 phase with perfect AFM order as the system is at half-filling.  The role played by $h_z$ here is similar to the sublattice potential that breaks the inversion symmetry in the noncentrosymmetric system  \cite{T.I.Vanhala,K.Jiang}.    
 
As SOC is a prerequisite for the emergence of TAFM, an important question is how the strength of SOC will affect the behavior of TAFM.  The upper inset of Fig. 4 shows the TAFM region in the parametric space of $(h_z, U)$ for certain SOC strength $t_s$.  When the SOC strength increases, a higher Hubbard $U$ is required to realize TAFM while the spanned region quickly shrinks.  This can be understood that when the in-plane hopping $t_s$ increases, it reduces the relative energy scale of $U$ effectively and delays the set-in of TAFM phases.  We also plot the TAFM region in the parametric space of $(t_s,U)$ for $h_z=0.1$, as in the lower inset of Fig. 4.  It shows when $t_s>0.8$, the TAFM phase will not appear anymore as the AFM order $\cal M_\perp$ induced by the strong SOC will be large enough to break the Dirac points and the topological configuration.  These results suggest that to observe the TAFM phase experimentally, the optimal parameter regime is a small external Zeeman field $h_z$ and a small ratio between Raman-assisted and usual hopping $|\frac{t_s}{t}|$, which is favorable in experiments as smaller $t_s$ implies weaker Raman-laser strengths and therefore lowers the heating rates \cite{H.Zhai}.

\section{Thermal Fluctuations}

In previous section, we illustrate the ground state phase diagrams and the properties of different phases at zero temperature.  Since thermal fluctuations usually play important roles in cold atom experiments, especially in Raman-assisted systems, we examine the finite-temperature effects in this section.  

\begin{figure}[h]
\includegraphics[width=8.8cm]{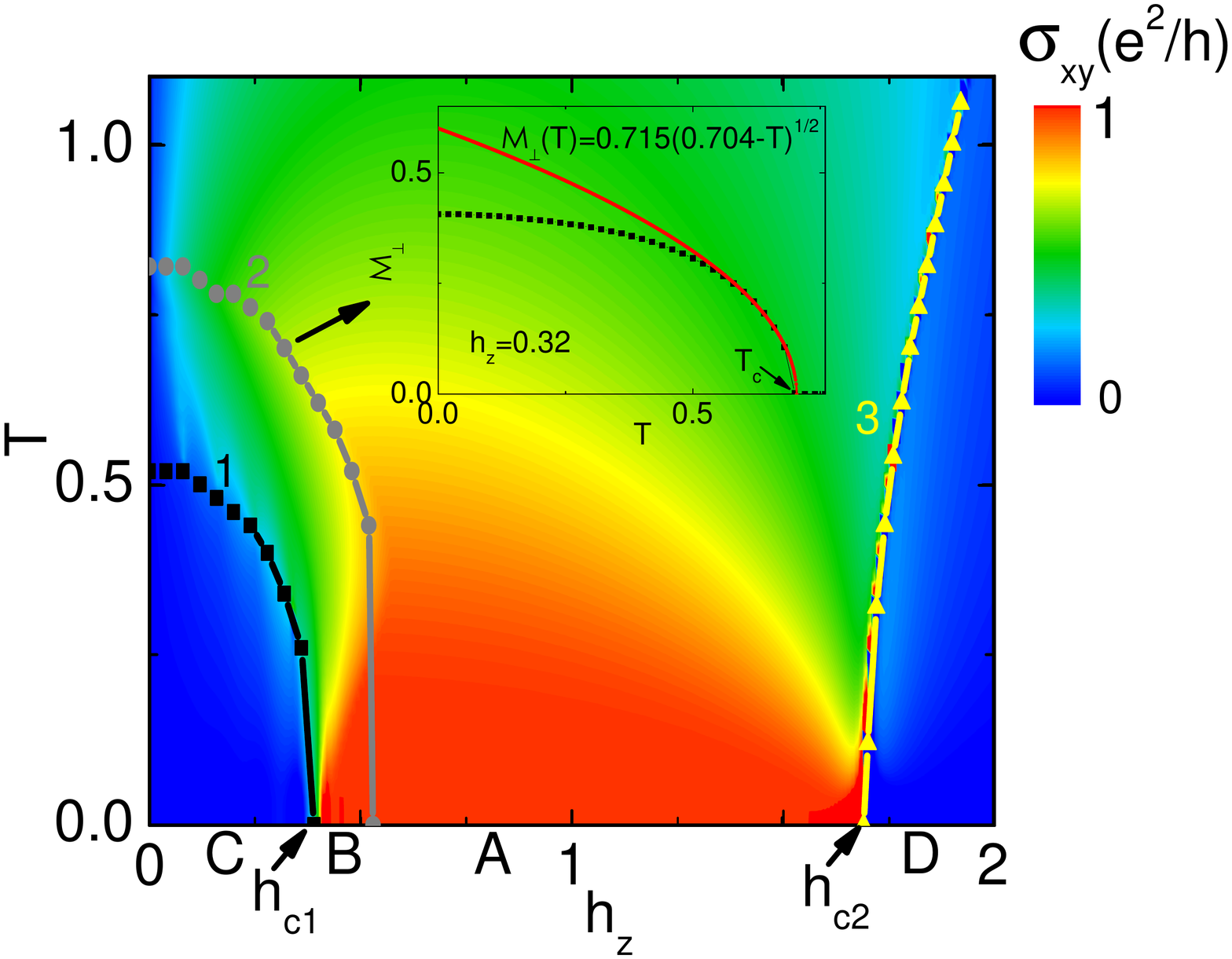}
\caption{(Color online) Finite-temperature phase diagram of the interacting system in the parametric space of $(h_z,T)$, with $t_s=0.48$ and $U=5$. The  Hall conductivity $\sigma_{xy}$ is shown as the contour plot with different colors.  The zero-temperature system includes the phases of C: normal AFM, B: TAFM, A: Chern insulator and D: trivial insulator.  The critical line 1 (black) and 3 (yellow) label the closure of the band gap when the SSB exists or not, respectively, while the critical line 2 (dark) labels the critical N$\acute{\text e}$el temperature, above which the SSB vanishes and ${\cal M}_\perp=0$.  The inset shows the dependence of $\cal M_\perp$ versus $T$ at $h_z=0.32$ and a fitting function ${\cal M}_\perp(T)$ around the critical point is also plot. \label{fig:finiteT}}
\end{figure}

The temperature has twofold effects in our system. First, in all phases, it creates excitations to higher Bloch bands with opposite Chern numbers, and therefore drives the Hall conductivity away from its quantized value at zero temperature. Secondly, the thermal fluctuations can shrink the regions of SSB phases in the phase diagram. To examine these effects more quantitatively, we compute the finite temperature Hall conductivity using the Kubo's formula \cite{D.J.Thouless,M.Tahir}.  The resulting phase diagram at finite temperature is shown in Fig.~\ref{fig:finiteT} in the parameter space $(h_z,T)$,  where the $T=0$ phase diagram was discussed previously in Fig.~\ref{fig:topo} (see the $U=5$ horizontal cut).  For the units $\frac{e^2}{h}$ of the Hall conductivity in Fig.~\ref{fig:finiteT}, $e$ is set by the effective coupling strength of the neutral atoms to synthetic external potential $\phi=-e\bm{E}_{\text{eff}}\cdot\bm{r}$, where $\bm{E}_{\text{eff}}$ is the effective electric field that can be synthesized by an off-center harmonic potential~\cite{M.Aidelsburger2}.

For the Hall conductivity $\sigma_{xy}$ in Fig. 5, it is clear that with the increasing of temperature, $\sigma_{xy}$ will be weaken, as expected.  When the SSB exists or not, the gap-closing of the bulk bands is denoted by the critical line 1 (black) or line 3 (yellow), respectively.  These critical lines also signal the Hall crossover from vanishing to a finite value.  When $h_z<h_{c1}$ ($h_z>h_{c2}$), the bulk gap will decrease (increase) with $h_z$, while the thermal fluctuations tend to close the gap, their combined effects lead to the decreasing (increasing) of the critical line 1 (3) with $h_z$. 

Fig.~\ref{fig:finiteT} also shows that the AFM order is suppressed by the thermal fluctuations at N$\acute{\text e}$el temperture $T_c$. The phase boundary is shown by line 2, above which ${\cal M}_\perp=0$ and the system reenters the rotational invariant phase.  With the increasing of $h_z$, the order parameter ${\cal M}_\perp$ decreases (see Fig. 3), which leads to the lowering of the critical $T_c$ to suppress the AFM order.  In the inset of Fig.~\ref{fig:finiteT} with $h_z=0.32$, we clearly see that the order parameter ${\cal M}_\perp$ changes continuously from a finite value to zero at $T_c$, pointing to the second-order phase transition.  The fitting function around the critical point can be given as ${\cal M}_\perp(T)=0.715(0.704-T)^{\frac{1}{2}}$, with the critical exponent $\beta=\frac{1}{2}$, which is in good consistent with the conventional scaling theory based on mean-field method.

\section{Discussions and Conclusions}

Compared with the previous works about the repulsive fermionic gas in Haldane model \cite{J.He1,W.Zheng}, the similar AFM phase with nontrivial topology was also found in a narrow region of the interacting phase diagram before entering the topologically trivial phase.  In this sense, we suggest the emergent of TAFM phase has certain universality for the repulsive fermion systems in Chern bands achieved through SOC.  In cold-atom system, as all parameters in the system can be precisely controlled, it provides a feasible platform to detect such a novel phase.  The topologically nontrival bands can be detected by measuring the orthogonal drifts of atoms after applying a constant force \cite{G.Jotzu,M.Aidelsburger2}.  While the AFM order of atoms can be measured from the Bragg scattering of light in cold-atom system \cite{T.A.Corcovilos,afm3}.  In experiment, the blue-detuned laser beams at wavelength $\lambda_L=767$nm are used to construct the lattice potential \cite{Z.Wu}.  If we choose $V_0=6E_R$, the tight-binding hopping integral corresponds to $t=0.111E_R$.  When taking $^{40}$K atom, the recoil energy $\frac{E_R}{\hbar}=2\pi\times 8.53$kHz and the temperature of $T=1$ corresponds to 45.3nK.  If we take $^6$Li atom, the recoil energy $\frac{E_R}{\hbar}=2\pi\times 56.86$kHz and $T=1$ corresponds to 301.7nK.  Both  temperatures are within the scope of present experimental detections \cite{Z.Wu}.  

To summary, we have studied the properties of interacting fermion atoms loaded in the optical lattice with Raman-assisted SOC and found the emergent of interaction-induced TAMF phase.  The further verifications of such topological magnetic-ordered state need more theoretical and experimental works in the future.

\acknowledgments

We would like to thank Biao Huang for helpful discussions.  This work was supported by China Scholarship Coucil (No. 201706795026) and NSF of Jiangsu Province of China (Grant No. BK20140129).

\end{document}